\documentclass[journal]{IEEEtran}
\pdfoutput=1
\ifCLASSINFOpdf
\else
   \usepackage[dvips]{graphicx}
\fi

\usepackage{graphicx}
\usepackage{subfigure}
\usepackage{float}
\usepackage{booktabs}
\usepackage{multirow}
\usepackage{verbatim}
\usepackage{amsmath}
\usepackage{bm}
\usepackage{hyperref}
\hypersetup{hidelinks,
	colorlinks=true,
	allcolors=black,
	pdfstartview=Fit,
	breaklinks=true}

\begin{document}

\title{Attention-Guided Generative Adversarial Network for Whisper to Normal Speech Conversion}

\author{Teng Gao, Jian Zhou, Huabin Wang, Liang Tao, and Hon Keung Kwan \IEEEmembership{}
\thanks{This research and development work was partly supported by Natural Science Fund Project of China under No.61301295, the Anhui Natural Science Fund Project (No.1708085MF151), and Anhui University Natural Science Research Project (KJ2018A0018).}
\thanks{Teng Gao, Jian Zhou, Huabin Wang and Liang Tao are with Key Laboratory of Intelligent Computing and Signal Processing, Ministry of Education, Anhui University, Hefei 230601 China (email: jzhou@ahu.edu.cn).}
\thanks{Hon Keung Kwan is with Department of Electrical and Computer Engineering, University of Windsor, Windsor, Ontario N9B 3P4, Canada.}}

\markboth{Journal of \LaTeX\ Class Files, Vol. xx, No. xx, August 2021}
{Shell \MakeLowercase{\textit{et al.}}: Bare Demo of IEEEtran.cls for IEEE Journals}
\maketitle

\begin{abstract}
Whispered speech is a special way of pronunciation without using vocal cord vibration. A whispered speech does not contain a fundamental frequency, and its energy is about 20dB lower than that of a normal speech. Converting a whispered speech into a normal speech can improve speech quality and intelligibility. In this paper, a novel attention-guided generative adversarial network model incorporating an autoencoder, a Siamese neural network, and an identity mapping loss function for whisper to normal speech conversion (AGAN-W2SC) is proposed. The proposed method avoids the challenge of estimating the fundamental frequency of the normal voiced speech converted from a whispered speech. Specifically, the proposed model is more amendable to practical applications because it does not need to align speech features for training. Experimental results demonstrate that the proposed AGAN-W2SC can obtain improved speech quality and intelligibility compared with dynamic-time-warping-based methods.
\end{abstract}


\IEEEpeerreviewmaketitle

\section{Introduction}

\IEEEPARstart{W}{hispered} speech is a special pronunciation style that has the following characteristics different from a normal voiced speech~\cite{A1}\cite{A2}: 1) The vocal cords do not vibrate when a person whispers, and its energy is about 20dB lower than that of a normal speech; 2) Since the vocal cords do not vibrate during pronunciation, the lungs need to exhale more airflow to stimulate the narrow semi-open glottis, resulting in a slower pronunciation and a longer sound length than that of a normal speech; 3) The formants of a whispered speech are shifted to a high frequency region, and the formants also have a wider bandwidth and a flatter frequency spectrum than those of a normal speech.

There are many practical applications for a whispered speech. For example, if a person's vocal cords, larynx, or other joints related to a speech production were damaged, he or she can only whisper. In such cases, it will be useful to improve the speech quality and/or intelligibility by an efficient whisper to normal speech conversion system/device.

Classical speech analysis-synthesis models such as LPC (Linear Prediction Coding), MELP (Mixed Excited Linear Prediction), CELP (Code Excited Linear Prediction), and STRAIGHT (Speech Transformation and Representation Using Adaptive Interpolation of Weighted Spectrum) are frequently used in traditional whisper to normal speech conversion~\cite{A3}\cite{A4}\cite{A5}. The advantage of adopting a speech analysis-synthesis model is that it does not need to conduct the time-domain waveform reconstruction from the converted speech acoustic features. However, an analysis-synthesis model requires fundamental frequency estimating from a whispered speech. Unfortunately, a whispered speech is a special speech signal without vocal cords vibration, resulting in the absence of a fundamental frequency. Currently, as far as we know, there is no efficient method to accurately estimate the fundamental frequency from a whispered speech~\cite{A6}\cite{A7}.

To overcome the difference in the pronunciation duration between a whispered speech and a normal speech, existing whisper to normal speech conversion methods use dynamic time warping (DTW) to align the acoustic features of a whispered speech with those of a normal speech. However, DTW does not consider the special pronunciation characteristics of a whispered speech. Consequently, some phonemes of a whisper fail to be converted to its normal couterpart, resulting unsmoothness in the converted normal speech~\cite{A8}.

To address the aforementioned limitations, we propose in this paper an attention-guided generative adversarial network incorporating an autoencoder, a Siamese neural network, and an identity mapping loss function for whisper to normal speech conversion (AGAN-W2SC). Specifically, an attention module is used in the proposed AGAN-W2SC model to process local pivotal features and assign the weight coefficients of each region adaptively to guide time alignment. Our core contributions are three folds: Firstly, only the mel-spectrogram is used and extracted respectively from a whispered speech and a normal speech, so tedious multi-parameter mapping can be avoided. Secondly, the model is trained at the frame level, where speech used to train is no longer limited to a specific length of time, making it more flexible and suitable for whisper to normal speech conversion with different time durations. Thirdly, the model can adaptively reweight each component in frequency domain by the proposed self-attention module, resulting in automatic time alignment between a whispered speech and a normal speech.

\section{Proposed AGAN-W2SC model}

\subsection{Generative adversarial network (GAN)}

Recently, the GAN has attracted much attention because of its ability in simulating data distribution. The GAN uses game theory to update network weights alternately between the generator and discriminator~\cite{A9}.

In this paper, the relationship between a source whispered speech and a target normal speech is modeled by an "encoder-decoder" block realized by two multi-layer convolutional neural networks to construct the generator part of the GAN, and a multi-layer convolutional network is used to realize the discriminator part of the GAN.

\begin{figure}[htb]
\centerline{\includegraphics[width=\columnwidth]{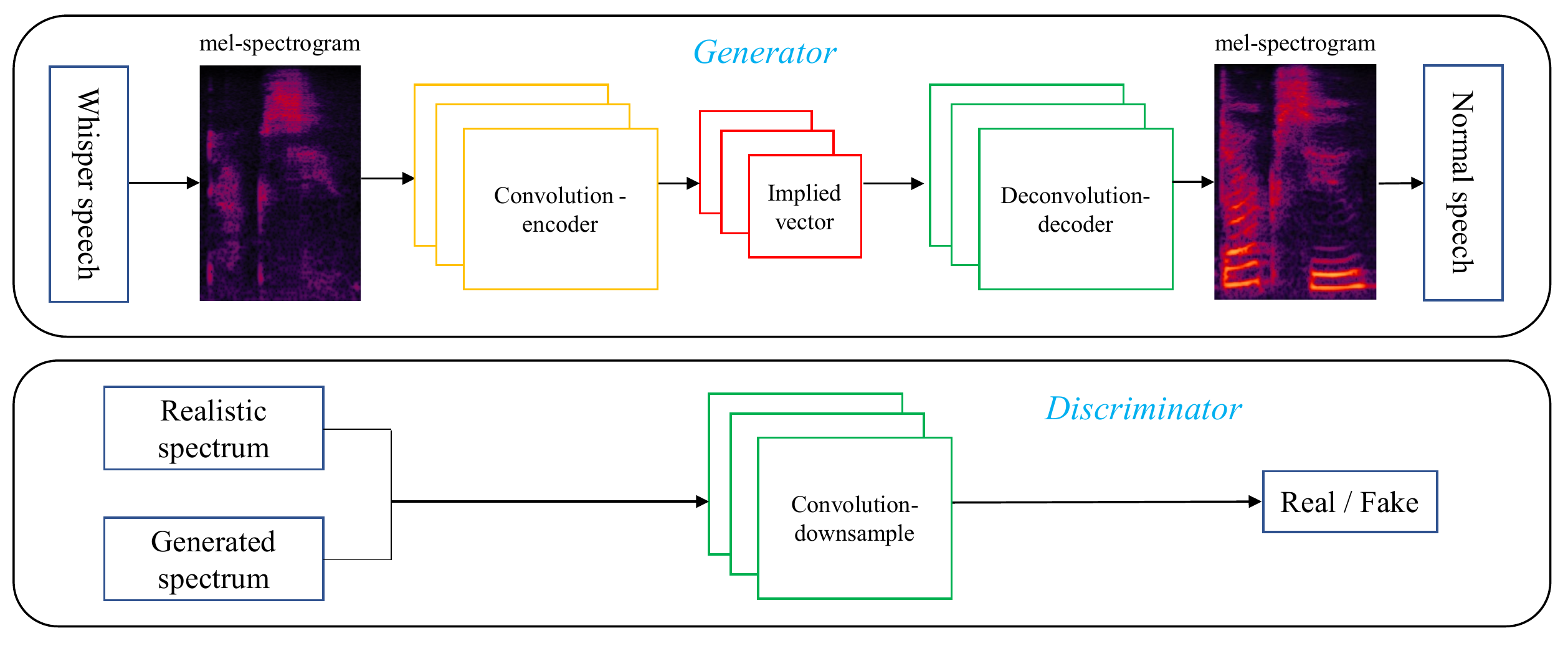}}
\caption{Proposed AGAN-W2SC model.}
\label{Fig.main1}
\end{figure}

Suppose \bm{$A$} denotes a source whispered speech spectrum feature vector, \bm{$B$} denotes a target normal speech spectrum feature vector, the generator and the discriminator in the GAN are denoted by $G$ and $D$ respectively, and let \bm{$B^{\prime}$} = $G$(\bm{$A$}) be the normal speech spectrum feature generated by the generator $G$. After feature extraction and dimension reduction of speech feature, a fully connected layer is adopted at the output of the multilayer convolutional network to output the tag value (Real/Fake), the proposed attention-guided generative adversarial network for whisper to normal speech conversion (AGAN-W2SC) is shown in Figure \ref{Fig.main1}. Table \ref{tab1} lists the configuration of generator with "encoder-decoder" architecture.

\begin{table}[ht] 
\centering
\caption{Configuration of generator}
\label{tab1}
\begin{tabular}{ccccccc}
\toprule
\multicolumn{2}{c}{Layer Name}& \multicolumn{1}{c}{Kernel}& \multicolumn{2}{c}{Input}& \multicolumn{2}{c}{Output} \\
\midrule
\multicolumn{2}{c}{Encoder-conv1}& \multicolumn{1}{c}{$3\times3$}& \multicolumn{2}{c}{$(128\times12)\times1$}& \multicolumn{2}{c}{$(128\times12)\times64$} \\
\multicolumn{2}{c}{Encoder-self-attention}& \multicolumn{1}{c}{}& \multicolumn{2}{c}{$(128\times12)\times64$}& \multicolumn{2}{c}{$(128\times12)\times64$} \\
\multicolumn{2}{c}{Encoder-padding}& \multicolumn{1}{c}{}& \multicolumn{2}{c}{$(128\times12)\times64$}& \multicolumn{2}{c}{$(128\times14)\times64$} \\
\multicolumn{2}{c}{Encoder-downsample}& \multicolumn{1}{c}{$128\times3$}& \multicolumn{2}{c}{$(128\times14)\times64$}& \multicolumn{2}{c}{$(1\times12)\times256$} \\
\multicolumn{2}{c}{Encoder-downsample}& \multicolumn{1}{c}{$1\times9$}& \multicolumn{2}{c}{$(1\times12)\times256$}& \multicolumn{2}{c}{$(1\times6)\times256$} \\
\multicolumn{2}{c}{Encoder-downsample}& \multicolumn{1}{c}{$1\times7$}& \multicolumn{2}{c}{$(1\times6)\times256$}& \multicolumn{2}{c}{$(128\times12)\times64$} \\
\multicolumn{2}{c}{\multirow{2}{*}{Decoder-upsample}}& \multicolumn{1}{c}{\multirow{2}{*}{$1\times7$}}& \multicolumn{2}{c}{$(1\times6)\times256$}& \multicolumn{2}{c}{\multirow{2}{*}{$(1\times6)\times512$}} \\
\multicolumn{2}{c}{}& \multicolumn{1}{c}{}& \multicolumn{2}{c}{$(1\times3)\times256$}& \multicolumn{2}{c}{} \\
\multicolumn{2}{c}{\multirow{2}{*}{Decoder-upsample}}& \multicolumn{1}{c}{\multirow{2}{*}{$1\times9$}}& \multicolumn{2}{c}{$(1\times6)\times512$}& \multicolumn{2}{c}{\multirow{2}{*}{$(1\times12)\times512$}} \\
\multicolumn{2}{c}{}& \multicolumn{1}{c}{}& \multicolumn{2}{c}{$(1\times12)\times256$}& \multicolumn{2}{c}{} \\
\multicolumn{2}{c}{Decoder-upsample}& \multicolumn{1}{c}{$128\times1$}& \multicolumn{2}{c}{$(1\times12)\times512$}& \multicolumn{2}{c}{$(128\times12)\times1$} \\
\bottomrule
\end{tabular}
\end{table}

\subsection{Self-attention module}

Different from conventional normal-to-normal voice conversions~\cite{A10}, the pronunciation speed of a whispered speech is different from a normal voiced couterpart. To tackle this issue, existing whispered speech conversion methods adopt a DTW algorithm to perform feature alignment. However, the converted normal speech obtained based on DTW exhibits a poor fluency. To address this problem, we configure a self-attention module~\cite{A11} in the generative adversarial network model to process the local pivotal features and assign the weight coefficients of each region adaptively to implicitly complete the time alignment.

\begin{figure}[htb]
\centerline{\includegraphics[width=\columnwidth]{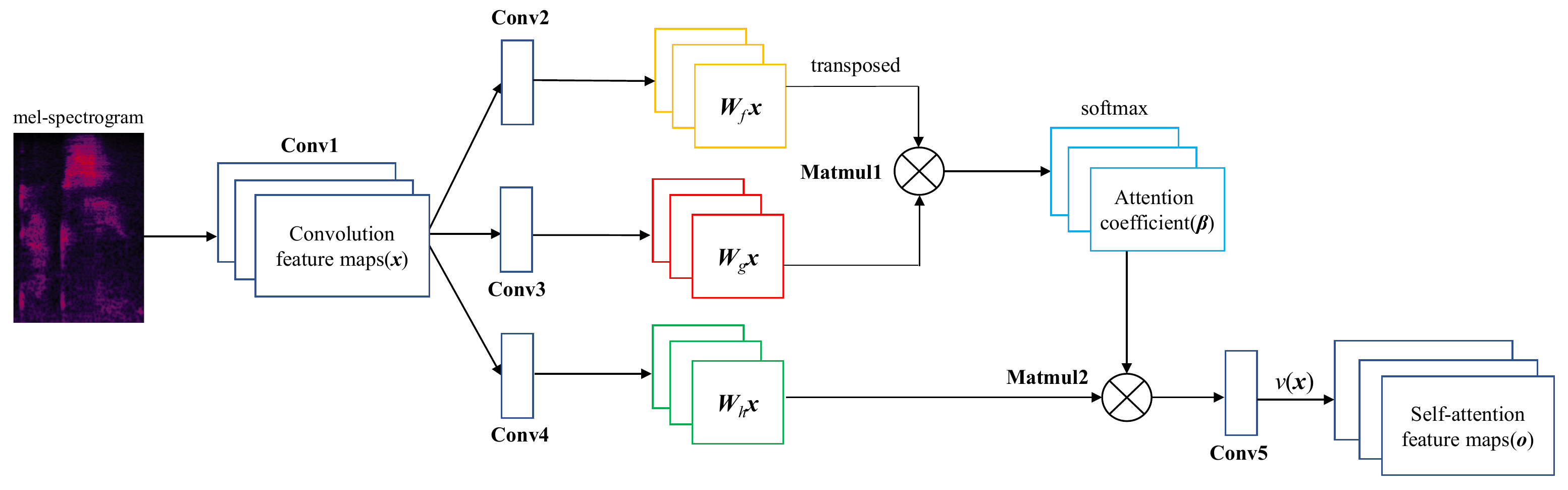}}
\centering \caption{Self-attention module with mel-spectrogram input.}
\label{Fig.main2}
\end{figure}

The self-attention mechanism adopted in this paper is shown in Figure \ref{Fig.main2}. The spectrum features after convolution from the previous hidden layer \bm{$x$} is first transformed into two feature spaces, i.e., \bm{$f$}, \bm{$g$} to calculate the attention, where \bm{$f$} = \bm{$W_{f}$}\bm{$x$}, \bm{$g$} = \bm{$W_{g}$}\bm{$x$}, we multiply \bm{$f^{T}$} by \bm{$g$} and do softmax to get the attention coefficient \bm{$\beta$}. At the same time, the hidden layer \bm{$x$} performs another convolution operation to get feature space \bm{$h$}, where \bm{$h$} = \bm{$W_{h}$}\bm{$x$}, and then we multiply \bm{$h$} and \bm{$\beta$} to get the initial attention feature maps. Finally, we perform convolution on initial attention feature maps to get self-attention feature maps \bm{$o$}. Note that the attention module is only used in the first encoder convolution layer of the generator and its configuration can be seen in Table \ref{tab2}.

\begin{table}[ht] 
\centering
\caption{Configuration of self-attention module}
\label{tab2}
\begin{tabular}{cccccccc}
\toprule
\multicolumn{2}{c}{Layer Name}& \multicolumn{2}{c}{Kernel}& \multicolumn{2}{c}{Input}& \multicolumn{2}{c}{Output} \\
\midrule
\multicolumn{2}{c}{Conv1}& \multicolumn{2}{c}{$3\times3$}& \multicolumn{2}{c}{$(128\times12)\times1$}& \multicolumn{2}{c}{$(128\times12)\times64$} \\
\multicolumn{2}{c}{Conv2}& \multicolumn{2}{c}{$1\times1$}& \multicolumn{2}{c}{$(128\times12)\times64$}& \multicolumn{2}{c}{$(128\times12)\times16$} \\
\multicolumn{2}{c}{Conv3}& \multicolumn{2}{c}{$1\times1$}& \multicolumn{2}{c}{$(128\times12)\times64$}& \multicolumn{2}{c}{$(128\times12)\times16$} \\
\multicolumn{2}{c}{\multirow{2}{*}{Matmul1}}& \multicolumn{2}{c}{}& \multicolumn{2}{c}{$(128\times12)\times16$}& \multicolumn{2}{c}{\multirow{2}{*}{$1536\times1536$}} \\
\multicolumn{2}{c}{}& \multicolumn{2}{c}{}& \multicolumn{2}{c}{$16\times(128\times12)$}& \multicolumn{2}{c}{} \\
\multicolumn{2}{c}{Softmax}& \multicolumn{2}{c}{}& \multicolumn{2}{c}{$1536\times1536$}& \multicolumn{2}{c}{$1536\times1536$} \\
\multicolumn{2}{c}{Conv4}& \multicolumn{2}{c}{$1\times1$}& \multicolumn{2}{c}{$(128\times12)\times64$}& \multicolumn{2}{c}{$(128\times12)\times128$} \\
\multicolumn{2}{c}{\multirow{2}{*}{Matmul2}}& \multicolumn{2}{c}{}& \multicolumn{2}{c}{$1536\times1536$}& \multicolumn{2}{c}{\multirow{2}{*}{$1536\times128$}} \\
\multicolumn{2}{c}{}& \multicolumn{2}{c}{}& \multicolumn{2}{c}{$(128\times12)\times128$}& \multicolumn{2}{c}{} \\
\multicolumn{2}{c}{Reshape}& \multicolumn{2}{c}{}& \multicolumn{2}{c}{$1536\times128$}& \multicolumn{2}{c}{$(128\times12)\times128$} \\
\multicolumn{2}{c}{Conv5}& \multicolumn{2}{c}{$1\times1$}& \multicolumn{2}{c}{$(128\times12)\times128$}& \multicolumn{2}{c}{$(128\times12)\times64$} \\
\bottomrule
\end{tabular}
\end{table}

\subsection{Frame-level feature segmentation and speech conversion}

Existing whisper to normal speech conversion methods conduct time-aligning for training corpus utterance by utterance. However, this complex time-aligning process cannot describe the implicit correlation between successive frames. To address this issue, sub-spectrums from mel-spectrogram are extracted every 12 frames and sent to the conversion model for batch training.

\begin{figure}[htb]
\centerline{\includegraphics[width=\columnwidth]{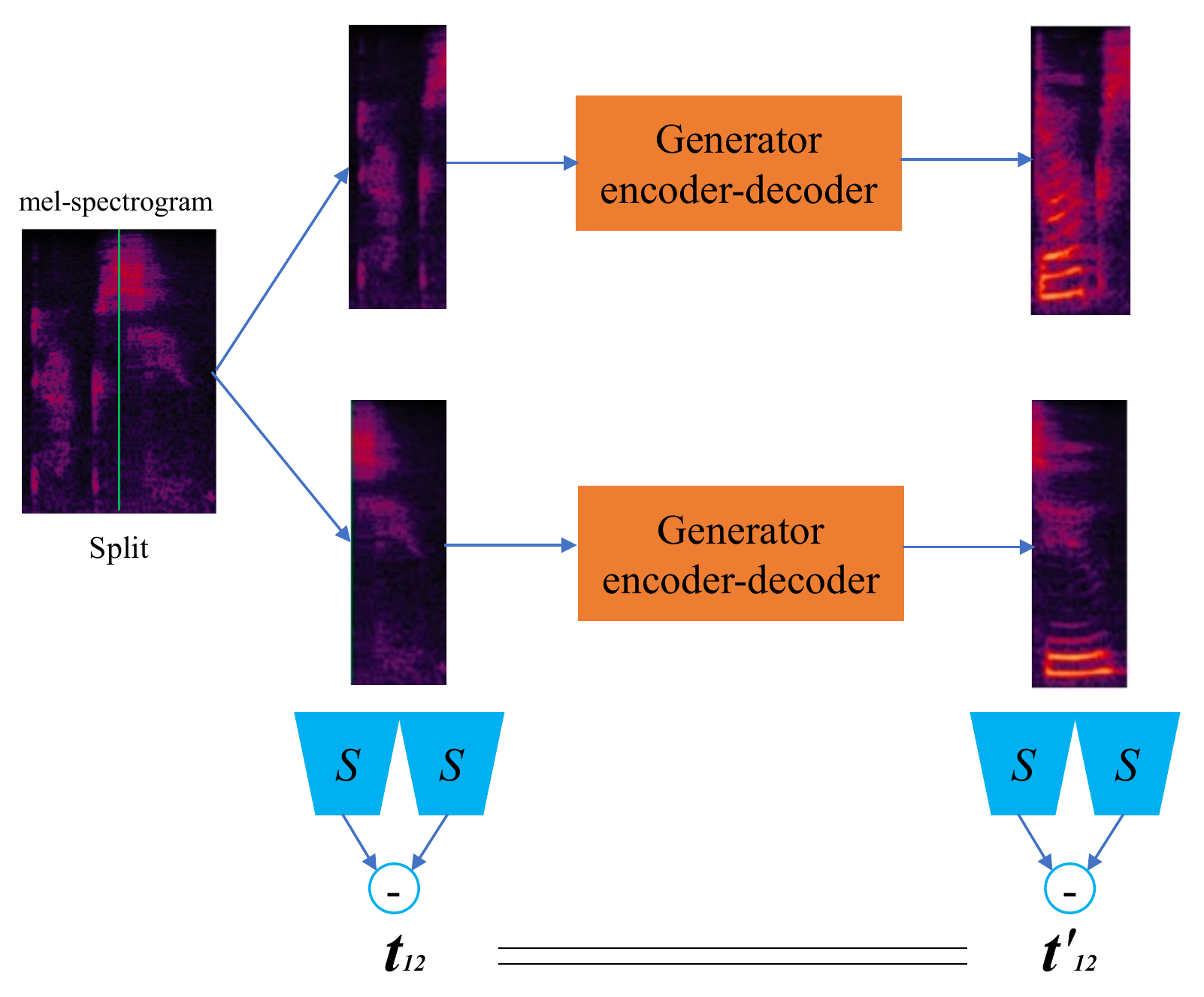}}
\caption{Siamese network used for transformation vector constraints.}
\label{Fig.main3}
\end{figure}

To keep the continuity of the generated sub-mel-spectrogram and make the synthesized normal speech sound more natural, we adopt a Siamese network~\cite{A12}\cite{A13} as shown in Figure \ref{Fig.main3} to constrain each pair of sub-mel-spectrograms to keep its first-order differential transformation features unchanged, ensuring the continuity of the spliced mel-spectrogram.

To keep the potential features of interest invariant, the first order difference vectors of the source whisper and the target normal speech are projected respectively to a low-dimensional hidden space with the constraint defined by Eq. (1).

\begin{eqnarray}
L_{(G,S)} = E_{(\bm{a_1},\bm{a_2})\sim \bm{A}} [\pi(\bm{t_{12}},\bm{t_{12}}^{\prime}) + \Vert \bm{t_{12}}-\bm{t_{12}}^{\prime} \Vert _2^2]        
\end{eqnarray}
where $$\bm{t_{12}} = S(\bm{a_1})-S(\bm{a_2})$$ 
      $$\bm{t_{12}}^{\prime} = S(G(\bm{a_1}))-S(G(\bm{a_2}))$$
      $$\pi(\bm{t_{12}},\bm{t_{12}}^{\prime}) = \frac{\bm{t_{12}} \cdot \bm{t_{12}}^{\prime}}{\Vert \bm{t_{12}} \Vert \Vert \bm{t_{12}}^{\prime} \Vert}$$

In Eqs. (1) and (2), $S$ denotes Siamese neural network, \bm{$a_1$} and \bm{$a_2$} are two different parts arbitrarily selected from the segmented mel-spectrogram, $\delta$ is a fixed value. With Eq. (2)~\cite{A14}\cite{A15}, $S$ encodes sub-mel-spectrogram so that in latent space each implied vector must be at least a predetermined fixed value $\delta$ apart from every other vector, avoiding the network from collapsing into a trivial function.

\begin{eqnarray}
L_{S} = E_{(\bm{a_1},\bm{a_2})\sim \bm{A}} \max(0,(\delta - \Vert \bm{t_{12}}\Vert _{2}))        
\end{eqnarray}

In addition, since traditional generative adversarial networks do not impose any constraint on a specific data information, the converted normal speech may lose some semantic information. To embedding more semantic information of the source whisper in the generated normal speech, the identity mapping loss~\cite{A16}\cite{A17} is also introduced in the generative adversarial network, and the generator $G$ is semantically constrained by Eq. (3).

\begin{eqnarray}
L_{(G,id)} = E_{\bm{b}\sim \bm{B}}[\Vert G(\bm{b}) - \bm{b} \Vert _1]
\end{eqnarray}

\section{Experiments}
\label{sec:experiment}

\subsection{Experimental configuration and evaluation metrics}

The whisper and normal-speech dataset adopted in~\cite{A18} is used in this paper. 800 pairs of parallel corpuses are randomly selected as the training set and the rest of 169 pairs as the testing set.

For training of AGAN-W2SC, 128x1 mel-spectrogram vector is extracted in each speech frame. Mel-spectrogram vectors of every successive 12 frames are concatenated and feeded to the model. The hinge loss defined by Eq. (4) and Eq. (5) is used for the adversarial training of generator and discriminator. Eq. (4) and Eq. (5) show that the discriminator $D$ learns how to better distinguish between real sample distribution \bm{$B$} and fake sample distribution $\bm{B}^{\prime}$ converted by the whispered speech \bm{$A$} via the generator, while the generator learns how to improve its mapping ability to confuse the discriminator. In this way, the fake sample distribution $\bm{B}^{\prime}$ generated by $G$ can approximate to the real sample distribution \bm{$B$} as much as possible.

$$L_{(D,adv)} = - E_{\bm{b}\sim \bm{B}}[\min(0,-1+D(\bm{b}))]$$
\begin{eqnarray}
~~~~~~~~~~~~~~~~~~~~~~~~- E_{\bm{a}\sim \bm{A}}[\min(0,-1-D(G(\bm{a})))] 
\end{eqnarray}
\begin{eqnarray}
L_{(G,adv)} = - E_{\bm{a}\sim \bm{A}}D(G(\bm{a})))
\end{eqnarray}

In the proposed AGAN-W2SC model, the spectrum normalization technique~\cite{A19} is used to improve the stability of the model by reducing the fluctuating of the output of the discriminator when the generated target mel-spectrogram exhibits a slight change.

For training the AGAN-W2SC model, one update step on the discriminator side for every 3 update steps on the generator side are used to speed up the model convergence. The learning rate is set between 0.0001 and 0.0002.

In the whisper to normal speech conversion stage, a time-domain waveform is produced from the output of the AGAN-W2SC model via the Griffin-Lim algorithm~\cite{A20}.

To evaluate the effectiveness of the proposed AGAN-W2SC model, the single-ended objective evaluation P.563~\cite{A21} is used. Baseline whisper to normal conversion models such as GMM~\cite{A26}, BLSTM~\cite{A27}, and CYCLE-GAN~\cite{A28} are also evaluated for comparison. The train set and test set of the baseline models are the same as the proposed AGAN-W2SC model. The extracted acoustic features used by the baseline models are aligned via DTW. A demo of the AGAN-W2SC model can be found at \url{https://mingze-sheep.github.io/b204_W2N.github.io/}.

\subsection{Experimental results and discussion}

P.563 evaluates speech quality without reference speech signal, it is especially suitable for the objective evaluation of whisper to normal speech conversion task. From Table \ref{tab3}, we can see that the proposed AGAN-W2SC model outperforms other DTW-based models.

\begin{table}[ht]
\centering
\caption{Objective evaluation results (single-ended indicator)}
\label{tab3}
%\begin{tabular}{lcl}
\begin{tabular}{p{70pt}p{30pt}}
\toprule
MODEL & P.563 \\
\midrule
WHISPER & 
1.009 \\
GMM & 
1.163 \\
BLSTM & 
1.247 \\
CYCLE-GAN & 
1.107 \\
AGAN-W2SC & 
\textbf{1.872} \\
NORMAL & 
2.667 \\
\bottomrule
\end{tabular}
\end{table}

\begin{figure}[htb]
\subfigure[]{
\begin{minipage}[c]{0.5\linewidth}
\centering
\includegraphics[width=0.99\linewidth]{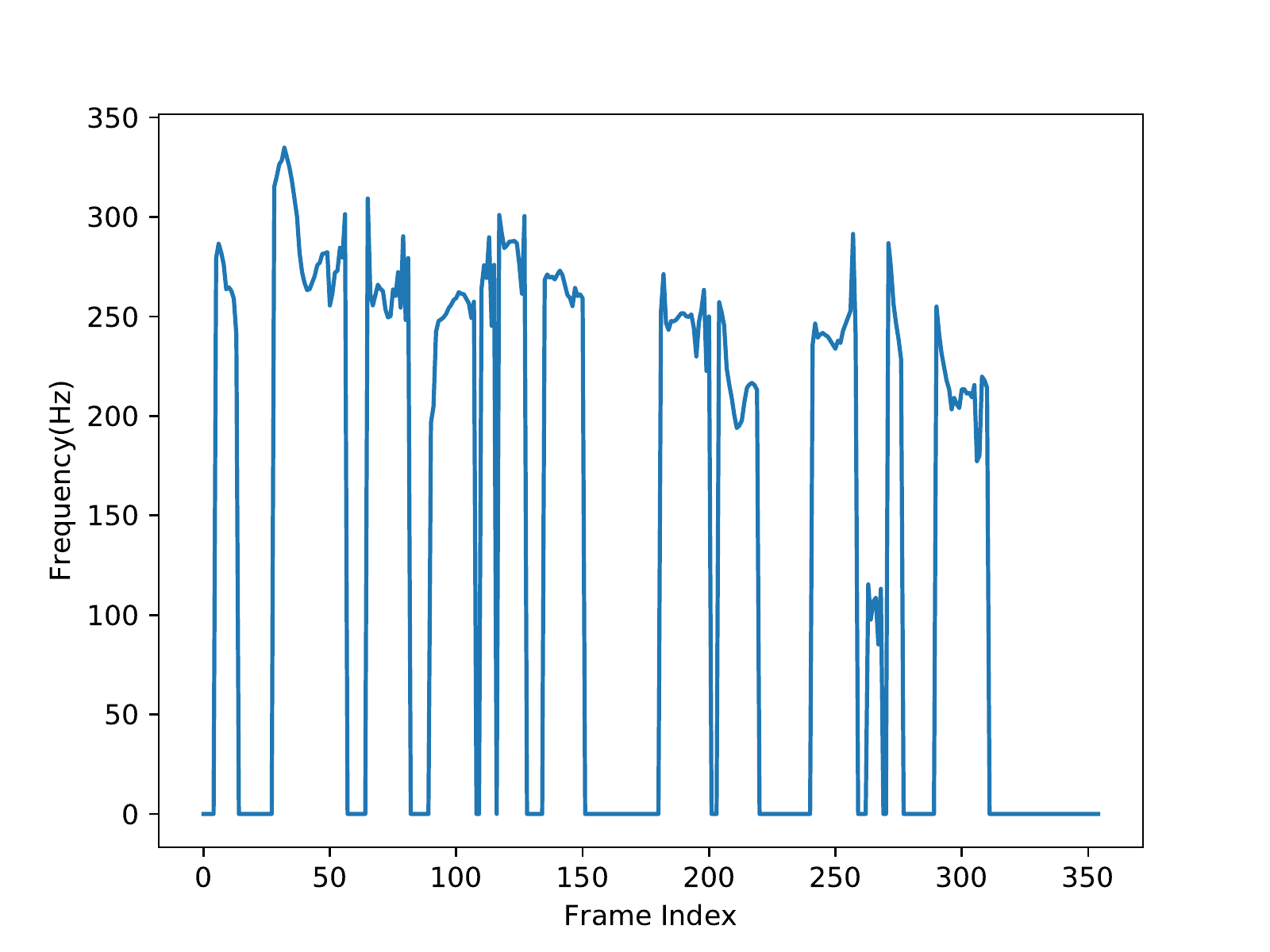}\vspace{4pt}
\end{minipage}}
\hspace{-5pt}
\subfigure[]{
\begin{minipage}[c]{0.5\linewidth}
\centering
\includegraphics[width=0.99\linewidth]{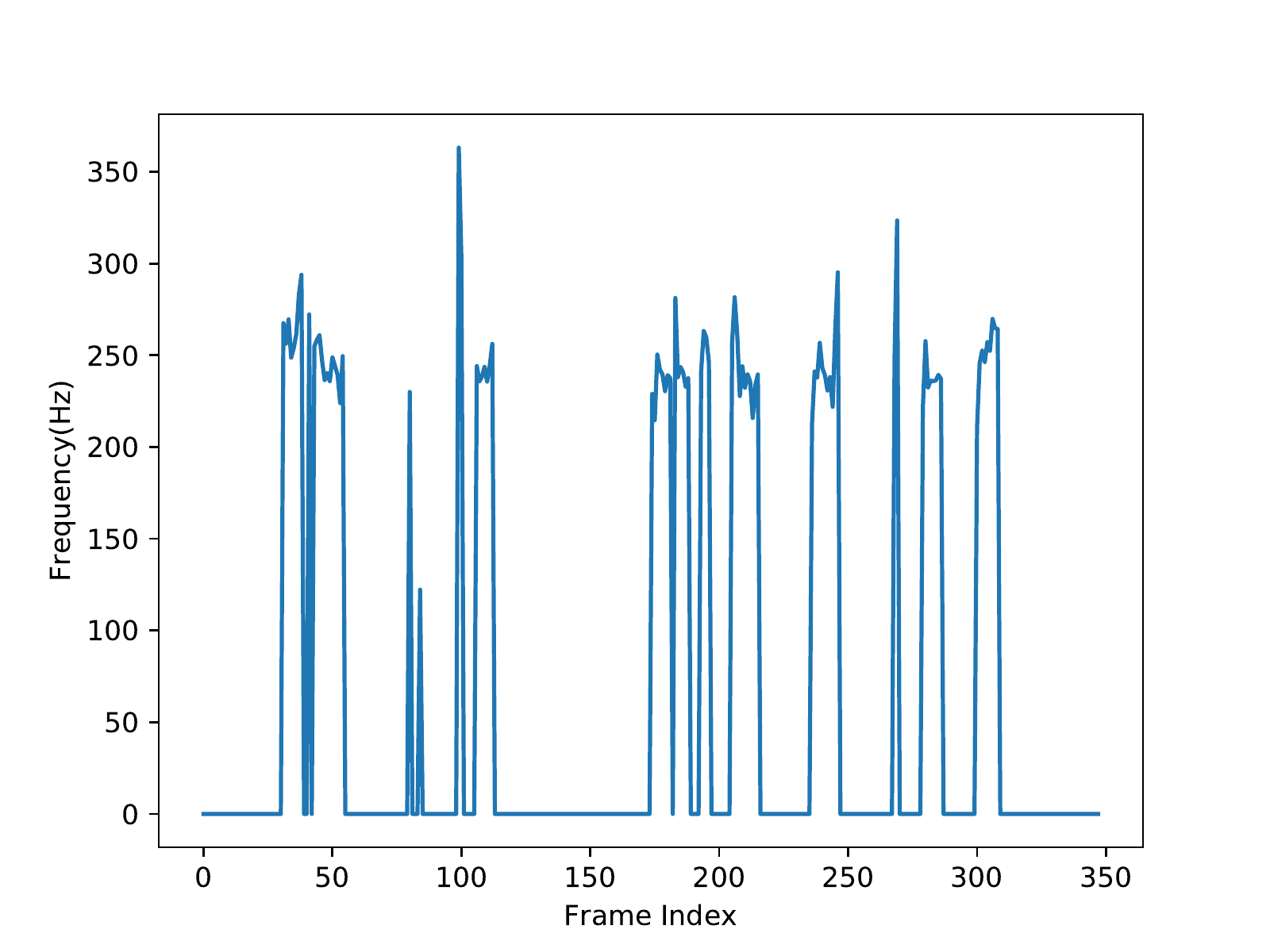}\vspace{4pt}
\end{minipage}}
\hspace{-50pt}
\subfigure[]{
\begin{minipage}[c]{0.5\linewidth}
\centering
\includegraphics[width=0.99\linewidth]{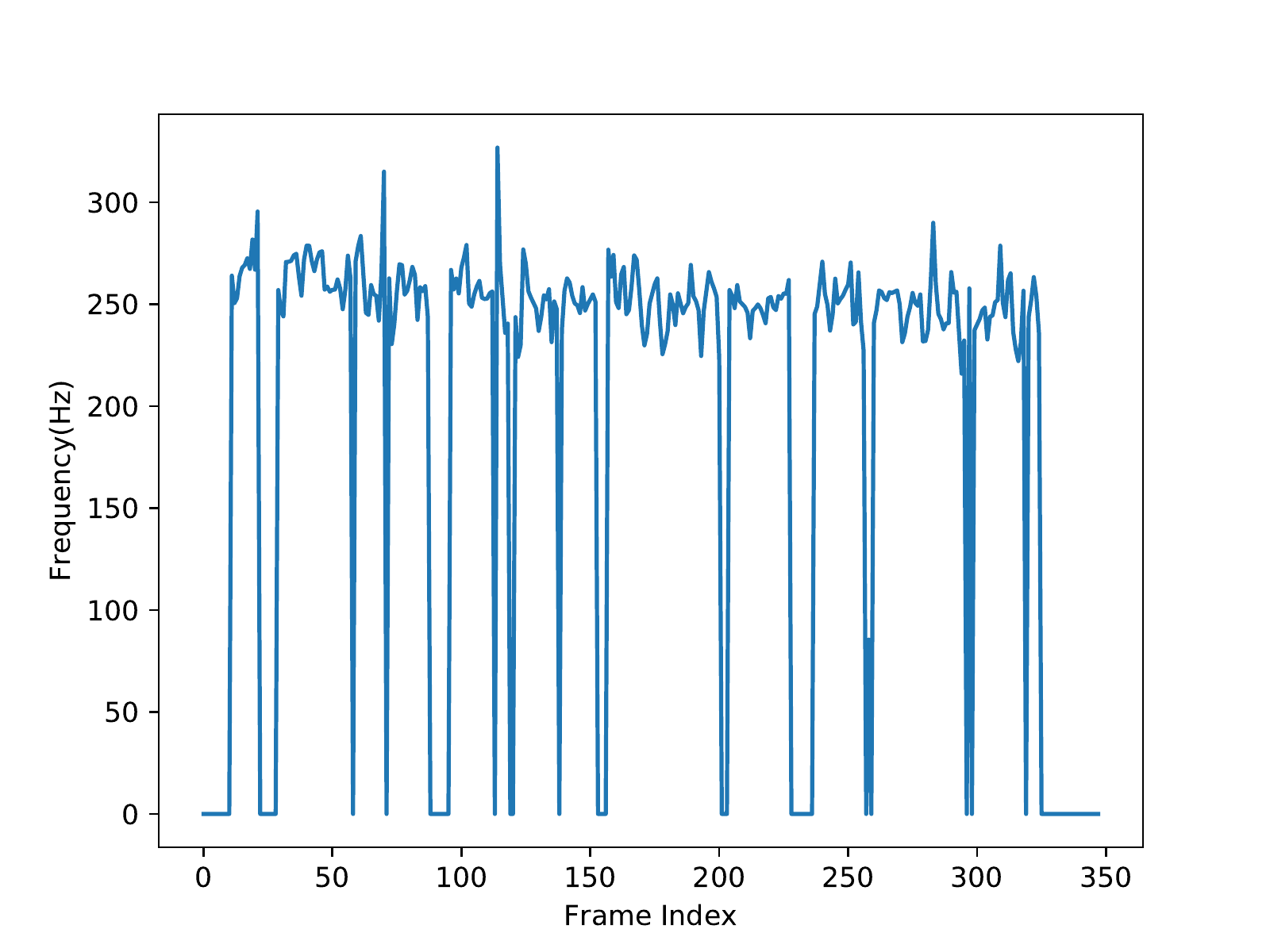}\vspace{4pt}
\end{minipage}
}\hspace{10pt}
\subfigure[]{
\begin{minipage}[c]{0.5\linewidth}
\centering
\includegraphics[width=0.99\linewidth]{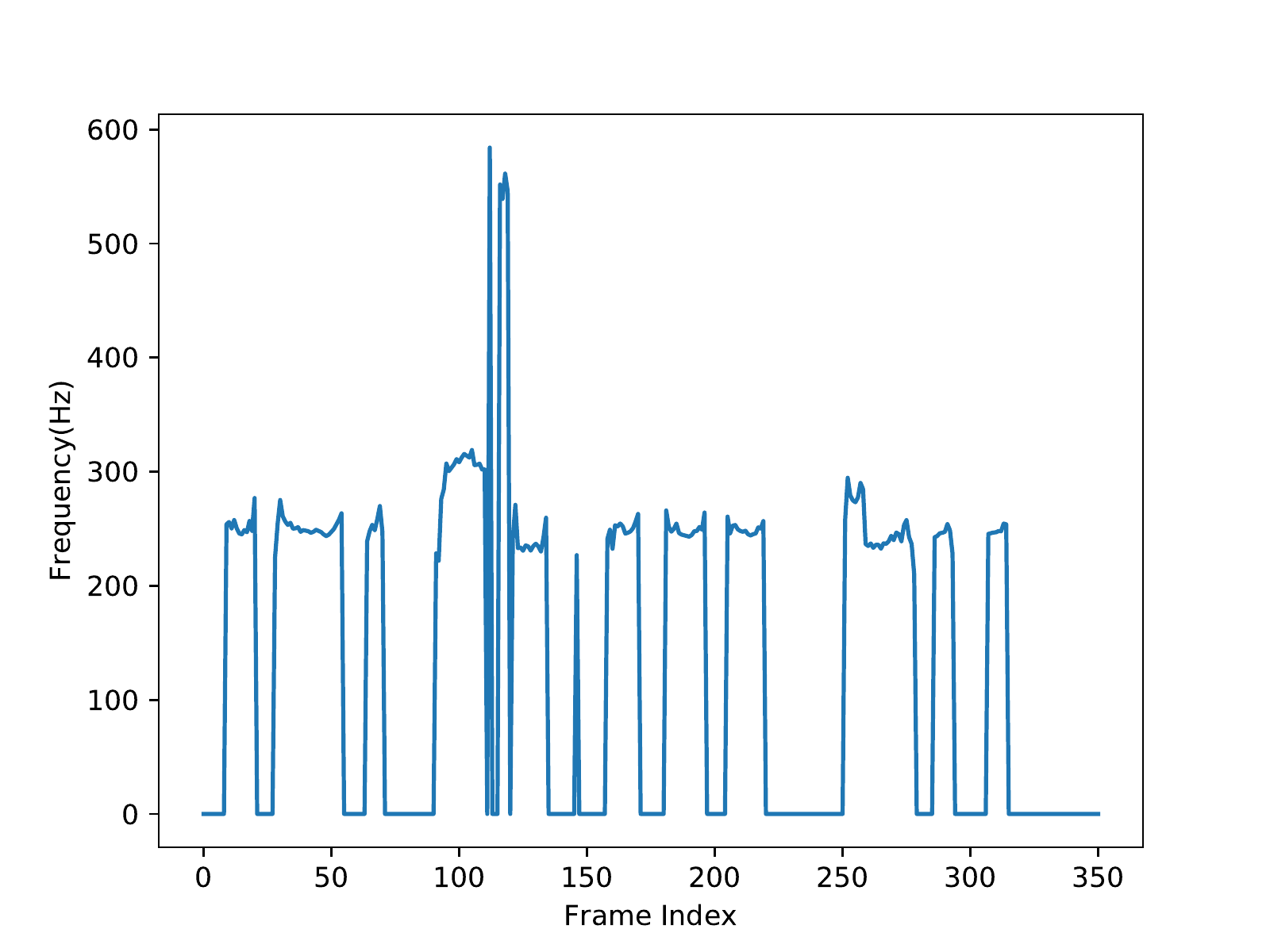}\vspace{4pt}
\end{minipage}
}
\hspace{-20pt}
\subfigure[]{
\begin{minipage}[c]{0.5\linewidth}
\centering
\includegraphics[width=0.99\linewidth]{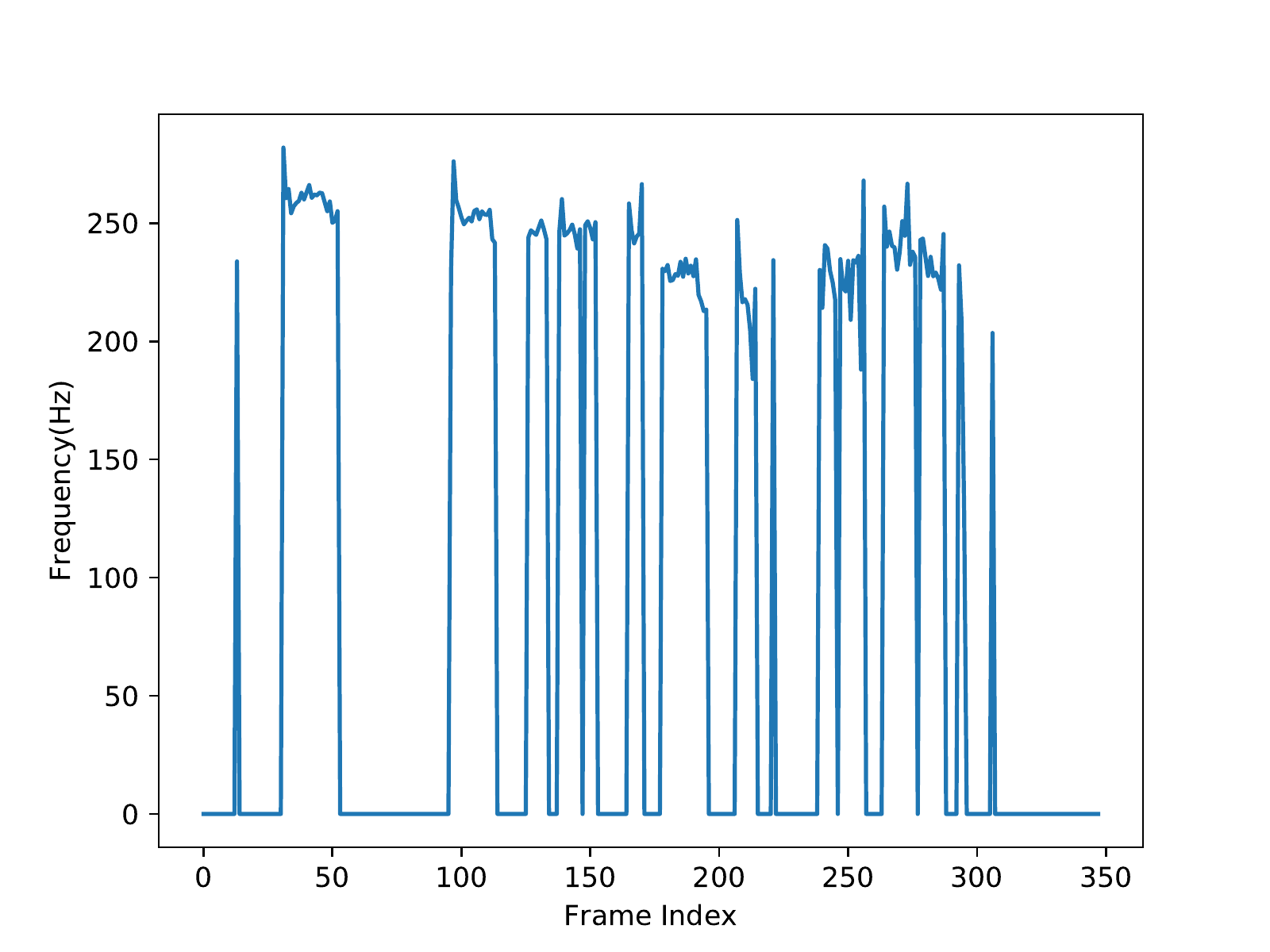}\vspace{4pt}
\end{minipage}
}
\caption{$F_0$ curve estimated by different whisper to normal speech conversion methods (test001.wav: The drug is also incorporated in supplements) (a) Reference normal speech; (b) GMM; (c) BLSTM; (d) Cycle-GAN; (e) AGAN-W2SC.}
\label{Fig.main4}
\end{figure}

Existing whisper to normal speech conversion methods aim to explicitly predict the missing fundamental frequency components from a whisper. To show the effectiveness of the proposed method for generating implicitly the pitch of the estimated normal speech from a whisper, we draw the $F_0$ curves of the reference normal speech and the converted speeches generated by different methods in Figure \ref{Fig.main4}. As shown in Figure \ref{Fig.main4}, the $F_0$ curve obtained by AGAN-W2SC exhibits a closer frequency contour with that of the reference normal speech, indicating that the model can effectively and competitively generate the fundamental frequency which is missed in a whispered speech.

We calculate the root mean square error (RMSE)~\cite{A29} of the fundamental frequency of the converted normal speech and the reference normal speech using 169 utterances. The RMSE equation is defined in Eq. (6), where $k$ is the total number of whisper-normal speech pairs after alignment, and $F_{0}^c$ and $F_{0}^t$ are the $F_0$ of the converted and the reference speech signals, respectively. There are two ways to calculate the RMSE value of the fundamental frequency: one is to directly calculate the $F_0$ of the speech aligned by DTW, which we call $F_0\_original$ in this paper; the other is by removing the silence frame after time alignment and then calculate it, which we call $F_0\_processed$ in this paper. 

\begin{eqnarray}
RMSE(F_0) = \sqrt{\sum_{i=1}^{k}{(F_{0_i}^c - F_{0_i}^t)^2}}   
\end{eqnarray}

\begin{figure}[htb]
\centerline{\includegraphics[width=\columnwidth]{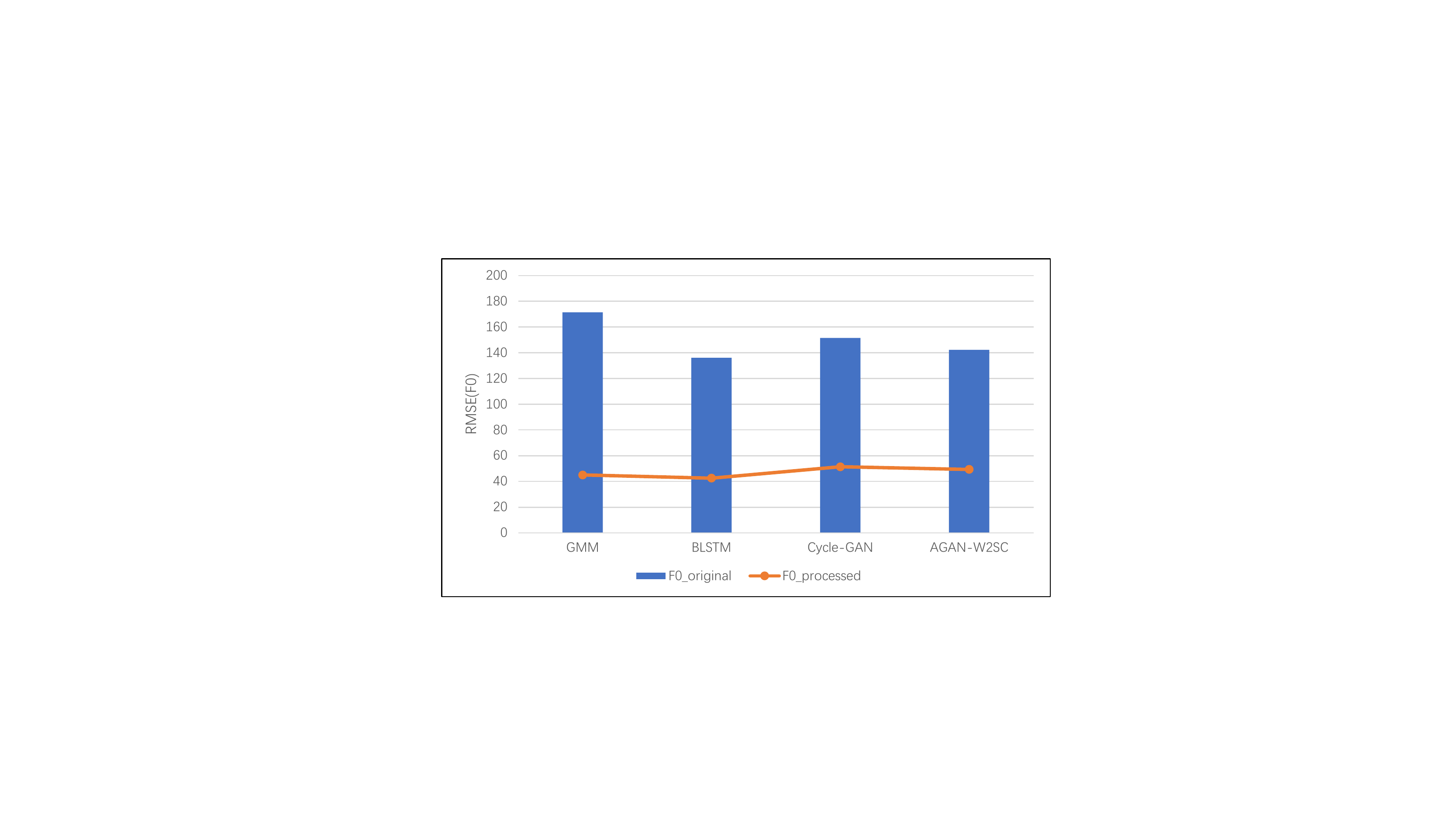}}
\caption{RMSE of estimated $F_0$ and ground truth $F_0$.}
\label{Fig.main5}
\end{figure}

As can be seen from Figure \ref{Fig.main5}, the proposed AGAN-W2SC has no special advantages over other models. This is due to the fundamental frequency of the speech is extracted frame by frame and then summed to get the RMSE value. The lowest RMSE value does not necessarily mean that every frame of speech is good, especially for some key frames. In addition, for calculating RMSE of $F_{0}$, it is necessary to align the converted speech and the reference normal speech to be of the same length in sample points, resulting in some distortions in speech quality. In fact, in terms of subjective auditory perception, AGAN-W2SC is superior to the other three baseline models. The two RMSE results shown in Figure \ref{Fig.main5} show that our proposed method can generate effective and competitive fundamental frequency components.

\section{Conclusion}

In this paper, a novel whisper to normal speech conversion model realized by an attention-guided generative adversarial network incorporating an autoencoder has been presented. Different from existing methods where speech corpus needs to be aligned by DTW before training, the proposed model is based on frame-level speech feature and conducts time-aligning adaptively through a self-attention mechanism. While the traditional methods explicitly predict the fundamental frequency of a target normal speech, the proposed method can implicitly generate the fundamental frequency from a whispered speech. Experimental results have indicated that the proposed method is effective and competitive for whisper to normal speech conversion.

%~\\
%~\\

%\section*{References}

\end{document}